\documentclass[preprint,amsmath,amssymb,aps,nofootinbib,showpacs]{revtex4}
\usepackage{graphicx}
\usepackage{bm}
\usepackage{subfigure}
\usepackage{amsmath}
\newcommand{\pr}{Phys. Rev.\ }

\newcommand{\jpa}{J. Phys. A\ }
\newcommand{\jpb}{J. Phys. B\ }

\newcommand{\etal}{{\em et al. }}
\newcommand{\e}{\mbox{e}}

\newcommand{\UQ}{School of Mathematics and Physics, University of Queensland, Brisbane, 
Queensland 4072, Australia.}

\begin{document}

\title{Entanglement properties of third harmonic generation via direct and cascaded parametric processes.}

\author{M.~K. Olsen}
\affiliation{\UQ}
\date{\today}

\begin{abstract}

We compare the bipartite entanglement and EPR-steering properties of the two different schemes which produce third harmonic optical fields from an input field at the fundamental frequency. The first scheme uses second harmonic cascaded with sum-frequency generation while the second uses triply degenerate four-wave mixing, also known as direct third harmonic generation. We examine and compare the two schemes in both the travelling wave and intra-cavity configurations. We find that both schemes produce continuous-variable bipartite entanglement and EPR-steering. The direct scheme produces a greater degree of EPR-steering while the cascaded scheme allows for greater flexibility in having three available bipartitions.

\end{abstract}


\maketitle

\section{Introduction}
\label{sec:intro}

The theoretical study of interactions of the electromagnetic field in nonlinear media goes back at least as far as 1962~\cite{Armstrong}, when Armstrong \etal published a classical treatment of the processes
of second- and third-harmonic generation, degenerate
and nondegenerate down-conversion, and four-wave mixing. Third-harmonic
generation (THG), wherein input fields at frequency $\omega$ produce
output fields at frequency $3\omega$ is a process that has been
observed experimentally in a number of different situations.
An early experiment~\cite{Tsang} produced both third and fifth harmonic
light at the interface of glass and liquids and it was
suggested that odd-multipole generation may be a widespread
phenomenon. THG has been observed in a number of
other situations, for example, in optical second harmonic generation cascaded with sum frequency generation (SFG)~\cite{Zhu}, in the interaction of laser light
with a nematic liquid-crystal cell\cite{Yelin}, and in the interaction
of pulsed light from an Nd:YAG (yttrium aluminum garnet)
laser with organic vapors~\cite{Ganeev1} and with polyimide films~\cite{Ganeev2}. 

There are at least two methods of producing optical fields where one is the third harmonic of another. The first is the method of direct third harmonic generation from a fundamental field, which is characterised as a triply degenerate four-wave mixing process. This method has been theoretically analysed by Gevorkyan \etal in the intracavity configuration~\cite{KVK3}, and by Olsen \etal in the travelling-wave configuration~\cite{THG}. The latter included a generalised Fokker-Planck equation with third-order partial derivatives for the positive-P function~\cite{P+}, which was then mapped onto stochastic differential equations(SDE) in an extended positive-P representation~\cite{P++}. The resulting SDEs were not very satisfactory, being prone to worse stability problems than those of the standard positive-P representation equations. The second method is to use cascaded second harmonic and sum frequency generation. Using this method, third harmonic generation was shown to be possible inside a quasi-periodic optical lattice by Zhu \etal in 1997~\cite{Zhu}.

The cascaded method has the advantage that it is amenable to a theoretical treatment using the positive-P method, as there are no terms in the Hamiltonian of higher than third order in creation and annihilation operators. The scheme has been analysed theoretically using this powerful method in both the travelling wave~\cite{Yu3,Yuthird} and intracavity~\cite{YuWang} configurations. These works have analysed the scheme in terms of both its mean-field and entanglement properties. The latter have been analysed using the method of symplectic eigenvalues~\cite{Serafini}, which is used to demonstrate that entanglement exists among the possible bipartitions of a multi-mode system. 

In this work we will extend previous research to investigate both configurations in terms of their entanglement properties, in both travelling wave and intracavity configurations. Rather than the symplectic eigenvalues, we will investigate the performance of these systems relative to the Duan-Simon~\cite{Duan,Simon} criteria for inseparability/entanglement and the Reid criteria~\cite{EPRMDR} for EPR-steering~\cite{EPR,Wisesteer}. These two criteria allow us to analyse in detail the entanglement present in each of the bipartitions of the cascaded system, and in the one possible bipartition for direct generation. We will use the positive-P representation for our numerical calculations. For the cascaded system it is an exact method, whereas for the direct generation scheme we will use a truncated positive-P approximation (TPPA) wherein the Fokker-Planck equation is truncated at the second-order derivatives before mapping onto stochastic differential equations.

\section{Hamiltonians and equations of motion}
\label{sec:Ham}

\subsection{Direct third harmonic generation}
\label{subsec:direct}

In direct third harmonic generation, an optical field at frequency $\omega_{a}$ acts to produce a field at $\omega_{b}$, where $\omega_{b}=3\omega_{a}$.
For the method, the interaction Hamiltonian is written as
\begin{equation}
{\cal H}_{I} = \frac{i\hbar\kappa}{3}\left(\hat{a}^{\dag\,3}\hat{b}-\hat{a}^{3}\hat{b}^{\dag}\right),
\label{eq:Hdirect}
\end{equation}
where $\hat{a}$ and $\hat{b}$ are the bosonic annihilation operators representing optical modes at frequencies $\omega_{a}$ and $\omega_{b}$ respectively. The constant $\kappa$ is related to the nonlinear polarisation of the medium. In the travelling wave configuration, the optical pumping of the nonlinear medium is related to the value of the fundamental mode at $t=0$, usually represented as the expectation value in a coherent state, although other pumping modes are possible, such as squeezed states of the field~\cite{Zela,Liz}, which does impact on the resulting quantum properties. We will use coherent state pumping here. When the nonlinear medium is held inside an optical cavity, the pumping is represented as
\begin{equation}
{\cal H}_{pd} = i\hbar\left(\epsilon_{d}\hat{a}^{\dag}-\epsilon_{d}^{\ast}\hat{a} \right),
\label{eq:Hpumpd}
\end{equation}
where the $\epsilon_{d}$ represent the coherent field incident on the cavity end mirror. The cavity damping is represented by the Liouvillian superoperator acting on the system density matrix,
\begin{equation}
{\cal L}_{d}\rho_{d} = \gamma_{a}\left(2\hat{a}\rho_{d}\hat{a}^{\dag}-\hat{a}^{\dag}\hat{a}\rho_{d}-\rho_{d}\hat{a}^{\dag}\hat{a} \right)+\gamma_{b}\left(2\hat{b}\rho_{d}\hat{b}^{\dag}-\hat{b}^{\dag}\hat{b}\rho_{d}-\rho_{d}\hat{b}^{\dag}\hat{b} \right),
\label{eq:Loueyd}
\end{equation}
where $\gamma_{a}$ and $\gamma_{b}$ are the cavity loss rates at the two frequencies. 

Following the standard methods~\cite{QNoise,DFW}, we can map the problem onto a Fokker-Planck equation for the Glauber-Sudarshan P-representation~\cite{Sud,RoyG}. A problem immediately becomes apparent in that this equation possesses partial derivatives of third-order, which means it cannot be mapped onto stochastic differential equations. For the purposes of this work, we will neglect these high-order derivatives in a manner analogous to that which leads to the often used truncated Wigner approximation~\cite{Robert}. This leads to another problem in that the diffusion matrix of the resulting Fokker-Planck equation is not positive-definite, which we solve by appealing to the positive-P representation~\cite{P+} for the approximated system. We will call the resulting representation the truncated postive-P approximation (TPPA). In the case of direct THG in the travelling wave configuration~\cite{THG}, this approximation was extremely accurate for the mean fields but tended to overstimate quantum correlations such as quadrature squeezing. However, the inaccuracy was merely quantitative, so that, given the unfavourable complications of the exact representation, we feel justified in using the approximate method here. The resulting It\^o~\cite{SMCrispin} equations for the intracavity scheme are
\begin{eqnarray}
\frac{d\alpha}{dt} &=& \epsilon-\gamma_{a}\alpha+\kappa\alpha^{+\,2}\beta+\sqrt{2\kappa\alpha^{+}\beta}\;\eta_{1}, \nonumber \\
\frac{d\alpha^{+}}{dt} &=& \epsilon^{\ast}-\gamma_{a}^{+}\alpha^{+}+\kappa\alpha^{2}\beta^{+}+\sqrt{2\kappa\alpha\beta^{+}}\;\eta_{2}, \nonumber \\ 
\frac{d\beta}{dt} &=& -\gamma_{b}\beta-\frac{\kappa}{3}\alpha^{3}, \nonumber \\
\frac{d\beta^{+}}{dt} &=& -\gamma_{b}\beta^{+}-\frac{\kappa}{3}\alpha^{+\,3},
\label{eq:SDEdirect}
\end{eqnarray}
where the $\eta_{j}$ are real Gaussian random variables with the properties $\overline{\eta_{j}(t)}=0$ and $\overline{\eta_{j}(t)\eta_{k}(t')}=\delta(t-t')\delta_{jk}$.
The complex variables $\alpha$ and $\beta$ correspond to the operators $\hat{a}$ and $\hat{b}$ in the sense that averages will converge approximately to expectation values of normally ordered operator moments such that $\overline{\alpha^{n}\alpha^{+\,m}}\rightarrow\langle\hat{a}^{\dag\,m}\hat{a}^{n}\rangle$. In general, $\alpha$ and $\alpha^{+}$ (same for $\beta$ and $\beta^{+}$) are not complex conjugates because of the independent noise terms, with this freedom allowing us to represent quantum evolution using classical c-number variables.
For systems which are represented accurately by positive-P stochastic differential equations, the equivalence between the complex variables and the operators is exact where the stochastic integration is stable. In the present case, using the TPPA, it is approximate, but has been shown to be accurate for intensities and qualitatively accurate in predicting squeezing~\cite{THG}.

\subsection{Cascaded third harmonic generation}
\label{subsec:cascade}

This process involves two parametric processes generated by a quasi-phase matched optical superlattice~\cite{Yuthird,Yu3,YuWang}. In the first, the fundamental mode at frequency $\omega_{0}$ acts via the nonlinearity represented by $\kappa_{1}$ to produce a harmonic mode at $\omega_{1}$, where $\omega_{1}=2\omega_{0}$. These two modes then interact via the $\kappa_{2}$ nonlinearity such that one photon from each combine to form a photon at $\omega_{2}$, with this being the third harmonic of $\omega_{0}$. This process thus cascades second harmonic generation with sum frequency generation~\cite{SFG} to indirectly produce polychromatic outputs, one of which is the third harmonic of another, from a single pump mode. Previous systems involving cascaded nonlinearities include non-degenerate downconversion as the first step~\cite{YuND,Clervie}, and multiple different modes interacting via twin nonlinearities~\cite{Ferraro,AxMuzz,OB}. Such systems have been shown to produce both bipartite and tripartite entanglement and EPR-steering~\cite{OCbitri,promiscuity}.

The interaction Hamiltonian for the cascaded process can be written as
\begin{equation}
{\cal H}_{I} = i\hbar\left(\kappa_{1}\hat{a}_{0}^{2}\hat{a}_{1}^{\dag}+\kappa_{2}\hat{a}_{0}\hat{a}_{1}\hat{a}_{2}^{\dag} \right) + h.c.,
\label{eq:Hcascade}
\end{equation}
while the cavity pumping Hamiltonian and the damping Liouvillian have the same basic forms as in the previous section. 
This system possesses a complete mapping onto the positive-P representation, resulting in the stochastic differential equations,
\begin{eqnarray}
\frac{d\alpha_{0}}{dt} &=& \epsilon-\gamma_{0}\alpha_{0}-2\kappa_{1}\alpha_{0}^{+}\alpha_{1}-\kappa_{2}\alpha_{1}^{+}\alpha_{2}+\sqrt{-2\kappa_{1}\alpha_{1}}\,\eta_{1}+\sqrt{-\kappa_{2}\alpha_{2}/2}\,(\eta_{3}+i\eta_{5}), \nonumber \\
\frac{d\alpha_{0}^{+}}{dt} &=& \epsilon^{\ast}-\gamma_{0}\alpha_{0}^{+}-2\kappa_{1}\alpha_{0}\alpha_{1}^{+}-\kappa_{2}\alpha_{1}\alpha_{2}^{+}+\sqrt{-2\kappa_{1}\alpha_{1}^{+}}\,\eta_{2}+\sqrt{-2\kappa_{2}\alpha_{2}^{+}/2}\,(\eta_{4}+i\eta_{6}), \nonumber \\
\frac{d\alpha_{1}}{dt} &=& -\gamma_{1}\alpha_{1}+\kappa_{1}\alpha_{0}^{2}-\kappa_{2}\alpha_{0}^{+}\alpha_{2}+\sqrt{-2\kappa_{2}\alpha_{2}/2}\,(\eta_{3}-i\eta_{5}), \nonumber \\
\frac{d\alpha_{1}^{+}}{dt} &=& -\gamma_{1}\alpha_{1}^{+}+\kappa_{1}\alpha_{0}^{+\,2}-\kappa_{2}\alpha_{0}\alpha_{2}^{+}+\sqrt{-2\kappa_{2}\alpha_{2}^{+}/2}\,(\eta_{4}-i\eta_{6}), \nonumber \\
\frac{d\alpha_{2}}{dt} &=& -\gamma_{2}\alpha_{2}+\kappa_{2}\alpha_{0}\alpha_{1}, \nonumber \\
\frac{d\alpha_{2}^{+}}{dt} &=& -\gamma_{2}\alpha_{2}^{+}+\kappa_{2}\alpha_{0}^{+}\alpha_{1}^{+},
\label{eq:PPcascade}
\end{eqnarray}
where the $(\alpha_{j},\alpha_{j}^{+})$ are the c-number variables corresponding to the operators $(\hat{a}_{j},\hat{a}_{j}^{\dag})$ in the same sense as in Eq.~\ref{eq:SDEdirect}.  Also, $\epsilon$ is the coherent pump amplitude, $\gamma_{j}$ is the cavity loss rate for mode $j$, and the $\eta_{j}$ are Gaussian random variables with the correlations $\overline{\eta_{j}(t)}=0$ and $\overline{\eta_{j}(t)\eta_{k}(t')}=\delta_{jk}\delta(t-t')$. Note that our noise terms are not exactly the same as those in Yu~\etal~\cite{Yu3,Yuthird,YuWang} and that this is due to the freedom available in the factorisation of the diffusion matrix of the Fokker-Planck equation for the positive-P function of the system. We have chosen to use real noises, which is a matter of taste since exactly the same physical system can be represented by either.

\begin{figure}[tbhp]
\includegraphics[width=0.75\columnwidth]{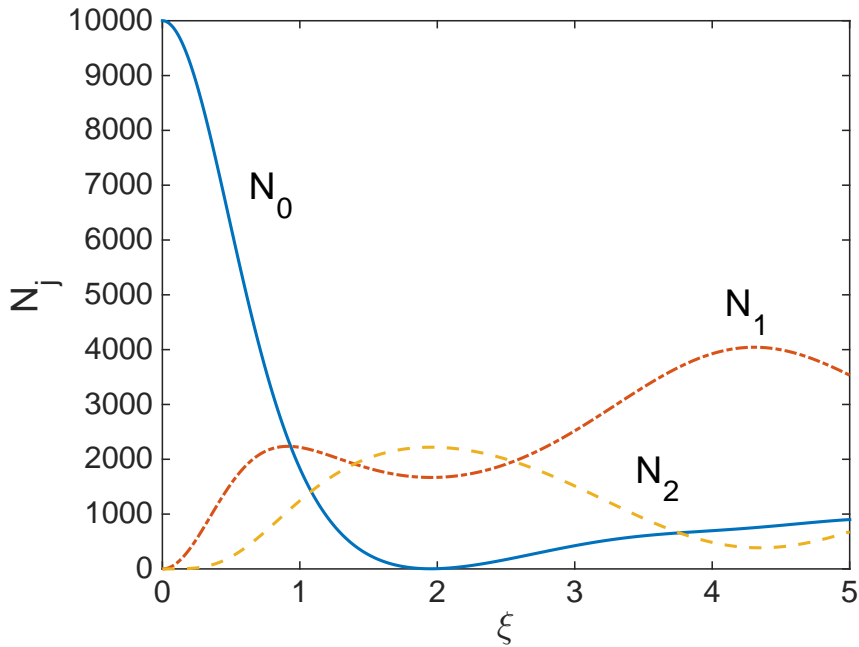}
\caption{(colour online) Mean field solutions for travelling wave third harmonic generation in the cascaded system, for $\kappa_{1}=10^{-2}$, $\kappa_{2}=1.5\kappa_{1}$, and $N_{0}(0)=10^{4}$. The dimensionless interaction time, $\xi$, is equal to $\kappa_{1} |\langle\alpha_{0}(0)\rangle |$. The positive-P equations were averaged over $3.6\times 10^{5}$ stochastic trajectories. Note that all quantities plotted in this and subsequent graphics are dimensionless.}
\label{fig:N0N1N2}
\end{figure}

As with the direct THG system, the travelling wave equations are found by removing the pumping and damping terms from Eq.~\ref{eq:PPcascade}. The solution of these equations for the field intensities is shown in Fig.~\ref{fig:N0N1N2}, for $\kappa_{1}=10^{-2}$, $\kappa_{2}=1.5\kappa_{1}$, and $N_{0}(0)=10^{4}$. We note here that these solutions are not consistent with those given by Yu \etal~\cite{Yuthird}, which reach a steady-state at $\xi\approx 6$, after which the modes at $\omega_{0}$ and $\omega_{2}$ are totally depleted, leaving only the second harmonic. As we know that the term $\kappa_{1}\hat{a}_{1}\hat{a}_{0}^{\dag\,2}$ in the Hamiltonian will act to downconvert the second harmonic with no other modes present, and this is in fact what is responsible for the revivals of the fundamental seen in travelling-wave SHG~\cite{tio}, we have confidence in our solutions. We note here that when we integrate the classical equations without any noise terms, we obtain the solutions given by Yu \etal 

\section{Quantum correlations in the travelling-wave configurations}
\label{sec:qtravel}

In this section we will give the results of stochastic integration of the travelling wave equations in the time domain for various quantum correlations. Using the quadrature definition
\begin{equation}
\hat{X}_{j}(\theta) = \hat{a}_{j}\e^{-i\theta}+\hat{a}_{j}^{\dag}\e^{i\theta},
\label{eq:Xdef}
\end{equation}
quadrature squeezing is found when the variance of $\hat{X}_{j}$ falls below one, for any $\theta$. When $\theta=0$, this is usually known as the X, or amplitude, quadrature, and when $\theta=\pi/2$, it is usually known as the Y, or phase, quadrature. In this work, all the correlations we present involve the X and Y quadratures, which would not be the case if our cavities were not resonant at all relevant frequencies~\cite{detune}.

\begin{figure}[tbhp]
\includegraphics[width=0.75\columnwidth]{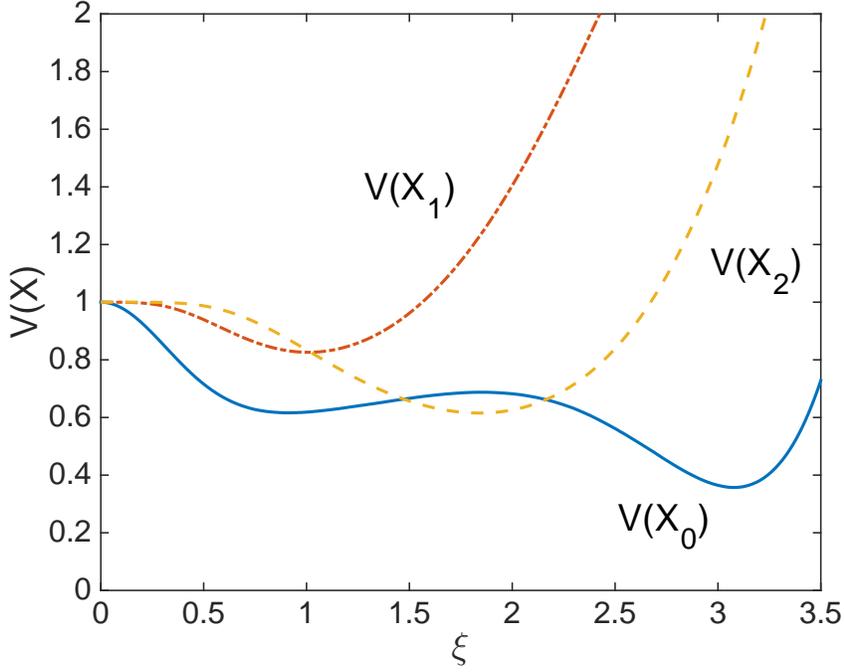}
\caption{(colour online) Variances in the X quadratures for travelling wave cascaded third harmonic generation, for the same parameters as in Fig.~\ref{fig:N0N1N2}. We see that the fundamental exhibits greater squeezing than the harmonic, and that both become anti-squeezed after some propagation length.}
\label{fig:VX0}
\end{figure}

The quadrature variances for the the direct configuration have previously been shown in Ref.~\cite{THG}, so we will show only those of Fig.~\ref{fig:VX0}, for the cascaded configuration. Some degree of squeezing is found in all the modes, although that in the first harmonic of the cascaded system is not pronounced. We found that it is possible to attain more squeezing of this mode by variation of the parameters $\kappa_{1}$ and $\kappa_{2}$, but at the expense of squeezing in the others.

Bipartite inseparability and entanglement are well characterised for continuous variable systems by either the Duan-Simon inequalities~\cite{Duan,Simon} or the Reid-EPR inequalities~\cite{EPR,EPRMDR}. The Duan-Simon inequalities, a violation of which is sufficient to prove bipartite entanglement for Gaussian systems, can be written as
\begin{equation}
V(\hat{X}_{i}\pm\hat{X}_{j})+V(\hat{Y}_{i}\mp\hat{Y}_{j})\geq 4.
\label{eq:DS}
\end{equation} 
We label these correlations $DS^{+}$ and $DS^{-}$, depending on whether the $X$ quadratures are added or subtracted.
The presence of the Einstein-Podolsky-Rosen (EPR) paradox~\cite{EPR} in bipartitions is signified by the well-known criteria developed by Reid~\cite{EPRMDR} in terms of inferred quadrature variances. The appropriate inequality is written as
\begin{equation}
V^{inf}(\hat{X}_{i})V^{inf}(\hat{Y}_{i})\geq 1,
\label{eq:EPRMDR}
\end{equation}
with violation of this signifying that the system demonstrates the EPR paradox. The inferred variances are defined as
\begin{eqnarray}
V^{inf}(\hat{X}_{i}) &=& V(\hat{X}_{i}) - \frac{\left[V(\hat{X}_{i},\hat{X}_{j}\right]^{2}}{V(\hat{X}_{j})},\nonumber\\
V^{inf}(\hat{Y}_{i}) &=& V(\hat{Y}_{i}) - \frac{\left[V(\hat{Y}_{i},\hat{Y}_{j}\right]^{2}}{V(\hat{Y}_{j})},
\label{eq:MDRinfs}
\end{eqnarray}
with the value of $\hat{X}_{i}$ being inferred from measurements of $\hat{X}_{j}$ (and similarly for $\hat{Y}_{i}$).
We immediately see that there is an implied asymmetry since we can equally define $V^{inf}(\hat{X}_{j})$. In some circumstances $i$ can be inferred from $j$, but not vice-versa, leading to a situation known as asymmetric steering~\cite{Wisesteer}. This was first predicted for Gaussian states in sum frequency generation~\cite{SFG} and later, also using Gaussian continuous-variable measurements, in the Kerr 
coupler~\cite{sapatona} and in intracavity second harmonic generation~\cite{meu}. It has also been demonstrated experimentally~\cite{Natexp}, again with Gaussian measurements. In what follows we will label the product $V^{inf}(\hat{X}_{i})V^{inf}(\hat{Y}_{i})$, inferred from $\hat{X}_{j}$ and $\hat{Y}_{j}$, as $EPR_{ij}$, noting that our analyses here deal with Gaussian measurements.

\begin{figure}[tbhp]
\includegraphics[width=0.75\columnwidth]{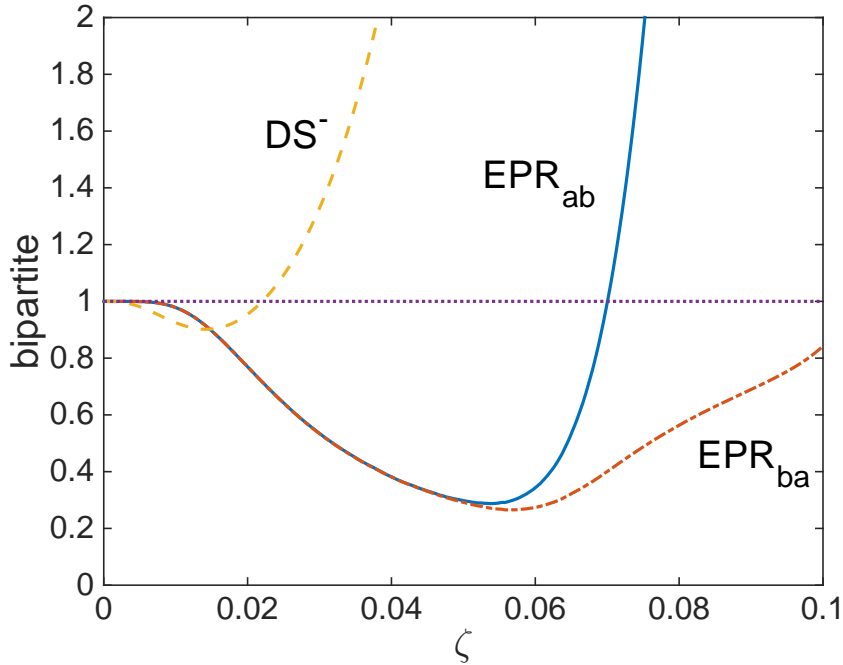}
\caption{(colour online) Bipartite correlations for travelling wave direct third harmonic generation, for $\kappa=10^{-3}$ and $N_{a}(0)=10^{4}$. The dimensionless interaction time, $\xi$, is equal to $\kappa |\langle\alpha(0)\rangle |$. The TPPA equations were averaged over $9\times 10^{7}$ stochastic trajectories.. Note that $DS^{-}$ has been divided by four, so as to have the same scaling as the Reid-EPR inequalities.}
\label{fig:THGbi}
\end{figure}

The results for the bipartite correlations in the direct configuration are shown in Fig.~\ref{fig:THGbi}. We see that there is bipartite entanglement between the modes at $\omega_{a}$ and $\omega_{b}$, and that the EPR inequalities are violated to a much greater extent than the $DS^{-}$. This is possible here because the system, being driven by a $\chi^{(3)}$ nonlinearity, produces outputs with non-Gaussian statistics, and a similar effect has been seen in atomic systems driven by s-wave collisions~\cite{noDS}. We find that the EPR-steering becomes asymmetric at $\zeta\approx 0.07$, which is around the time where upconversion plateaus and the fundamental begins to revive. After this time, measurements of the high frequency mode cannot be used to steer the fundamental, although the fundamental can still be used to steer the high frequency mode. We note here that these results are only qualitatively accurate, with there being two reasons for this. One is that we have used the TPPA, which is known to slightly overestimate squeezing in this system, and the other is that our analysis only includes unitary Hamiltonian dynamics, not including such effects as dispersion in the nonlinear medium. There are no possible tripartite entanglement correlations for this system, since there are only two distinguishable modes present.

\begin{figure}[tbhp]
\includegraphics[width=0.75\columnwidth]{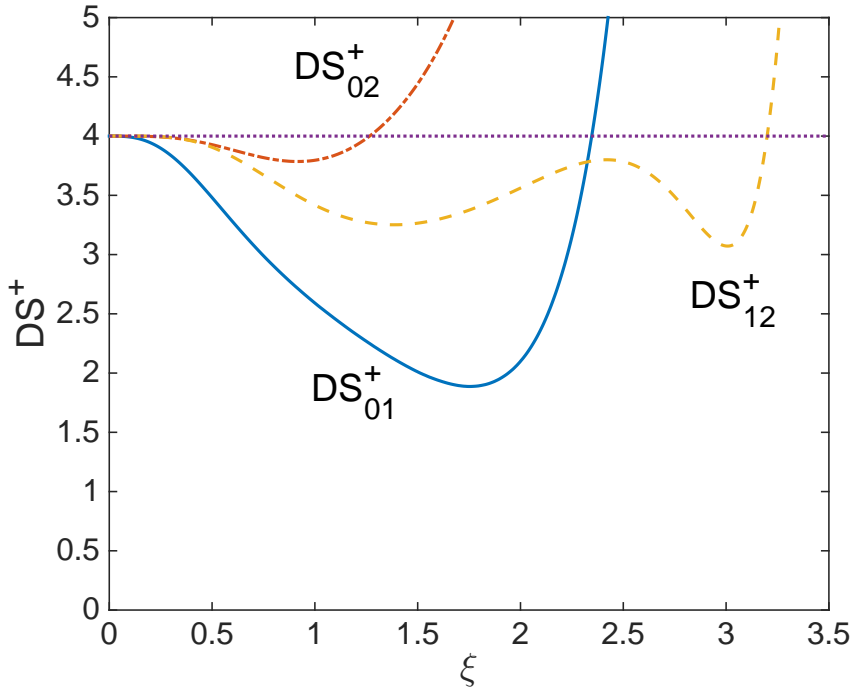}
\caption{(colour online) Duan-Simon correlations for travelling wave cascaded third harmonic generation, for the same parameters as in Fig.~\ref{fig:N0N1N2}.}
\label{fig:DScascade}
\end{figure}

In the cascaded configuration we have three possible bipartitions of the system and the symplectic eigenvalue analysis given by Yu \etal shows that at least some of these will exhibit entanglement, without specifying which pairs~\cite{Yuthird}. As shown in Fig.~\ref{fig:DScascade}, entanglement is found in all three bipartitions over some finite interaction times, using the Duan-Simon inequalities. In this case, the inequalities are violated to a greater extent between adjacent modes than they are between the fundamental and the third harmonic. This indicates that the direct method may be more efficient for the production of entangled states spanning the whole of the frequency range. We also looked for tripartite entanglement in this system, using the van Loock-Furusawa inequalities~\cite{vLF}. Over an extensive parameter range using different input amplitudes and ratios of $\kappa_{1}/\kappa_{2}$, we found no evidence of tripartite entanglement. The cascaded down conversion and sum frequency generation system investigated by Pennarun \etal does possess tripartite entanglement~\cite{Clervie} and we note here that absence of evidence is not the same as evidence of absence. In this context, we also note that an earlier analysis of sum frequency generation had found no evidence of tripartite entanglement over a broad parameter range~\cite{SFG}, whereas a more recent analysis has found that it is actually present for other parameters~\cite{SFGChina}. Therefore we can not rule out its presence in this cascaded system, although the three-colour entanglement mentioned in previous analyses should not be interpretated as tripartite entanglement, but rather as entanglement being present in bipartitions of modes at three different frequencies.

\begin{figure}[tbhp]
\includegraphics[width=0.75\columnwidth]{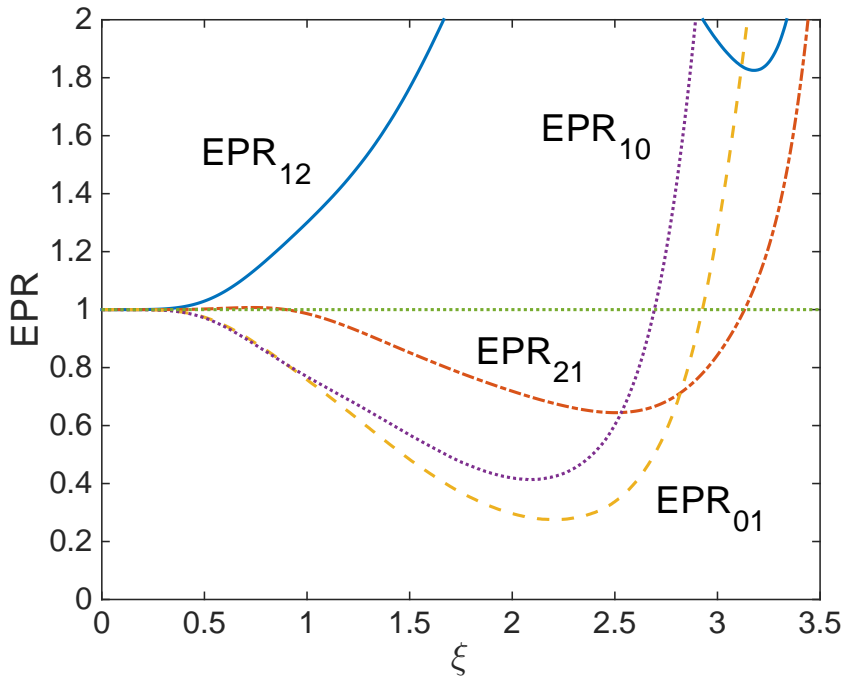}
\caption{(colour online) Reid EPR correlations for travelling wave cascaded third harmonic generation, for the same parameters as in Fig.~\ref{fig:N0N1N2}.}
\label{fig:EPRcascade}
\end{figure}

The bipartite EPR correlations which exhibit a significant degree of violation of the Reid inequalities for the cascaded system are shown in Fig.~\ref{fig:EPRcascade}. Those for the bipartition of $0$ and $2$ are not shown since only $EPR_{20}$ violates the inequality, and then by an insignificant amount, never dropping below a value of $0.99$. We see that the largest violation of the inequalities is found with the fundamental and the second harmonic, while the second and third harmonics show asymmetric steering for all interaction times at which EPR-steering is found for this bipartition. In the next section we will examine how the various configurations are changed when the interacting media are contained inside resonant optical cavities.

\section{Steady-state correlations for intracavity configurations}
\label{sec:cavidade}

When nonlinear optical media are held inside a pumped optical cavity, the accessible observables are usually the output spectral correlations, which are measurable using homodyne techniques~\cite{mjc}. These are readily calculated in the steady-state by treating the system as an Ornstein-Uhlenbeck process~\cite{SMCrispin}. In order to do this, we begin by expanding the positive-P variables into their steady-state expectation values plus delta-correlated Gaussian fluctuation terms, e.g.
\begin{equation}
\alpha_{ss} \rightarrow \langle\hat{a}\rangle_{ss}+\delta\alpha.
\label{eq:fluctuate}
\end{equation}
Given that we can calculate the $\langle\hat{a}\rangle_{ss}$, we may then write the equations of motion for the fluctuation terms. The resulting equations are written for the vector of fluctuation terms as
\begin{equation}
\frac{d}{dt}\delta\vec{\alpha} = -A\delta\vec{\alpha}+Bd\vec{W},
\label{eq:OEeqn}
\end{equation}
where $A$ is the drift matrix containing the steady-state solution, $B$ is found from the factorisation of the drift matrix of the original Fokker-Planck equation, $D=BB^{T}$, with the steady-state values substituted in, and $d\vec{W}$ is a vector of Wiener increments. As long as the matrix $A$ has no eigenvalues with negative real parts, this method may be used to calculate the intracavity spectra via
\begin{equation}
S(\omega) = (A+i\omega)^{-1}D(A^{\mbox{\small{T}}}-i\omega)^{-1},
\label{eq:Sout}
\end{equation}
from which the output spectra are calculated using the standard input-output relations~\cite{mjc}.

\subsection{Intracavity direct conversion}
\label{subsec:cavdirect}
    
The drift matrix for the fluctuations of this system is written as
\begin{equation}
A_{d} = 
\begin{bmatrix}
\gamma_{a} & -2\kappa\alpha^{\ast}\beta & -\kappa\alpha^{\ast\,2} & 0 \\
-2\kappa\alpha\beta^{\ast} & \gamma_{a} & 0 & -\kappa\alpha^{2} \\
\kappa\alpha^{2} & 0 & \gamma_{b} & 0 \\
0 & \kappa\alpha^{\ast\,2} & 0 & \gamma_{b}
\end{bmatrix},
\label{eq:Admat}
\end{equation}
where we have used the notation $\langle\hat{a}\rangle_{ss},(\langle\hat{b}\rangle_{ss})=\alpha,(\beta)$. The matrix $D_{d}$ has $[\sqrt{2\kappa\alpha^{\ast}\beta},\sqrt{2\kappa\alpha\beta^{\ast}} ,0,0]$ on the diagonal, with all other elements being zero, and the subscript signifies direct third harmonic generation.

\begin{figure}[tbhp]
\includegraphics[width=0.75\columnwidth]{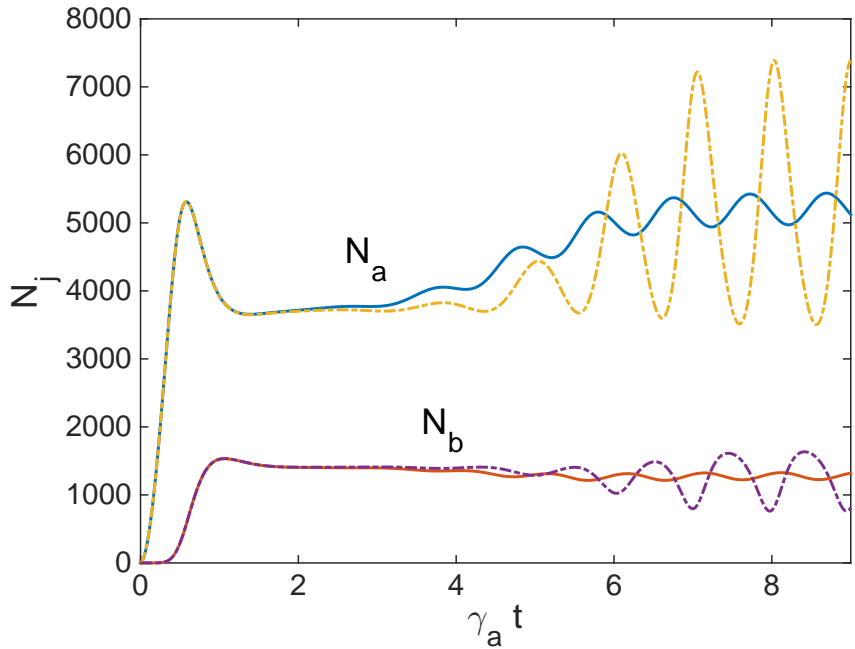}
\caption{(colour online) The classical (dash-dotted lines) and TPPA (solid lines) mean field solutions for intracavity direct THG. Parameters are $\gamma_{a}=1$, $\gamma_{b}=2\gamma_{a}$, $\kappa=10^{-3}$, and $\epsilon=200$, with initial conditions $\alpha(0) = 1+i$ and $\beta(0)=0$. The critical pump amplitude is $\epsilon_{c}\approx 137$.}
\label{fig:THGpulse}
\end{figure}

In this configuration, Gevorkyan \etal have shown that the matrix $A_{d}$ will have eigenvalues with negative real part for
\begin{equation}
\epsilon \geq\epsilon_{c} = \frac{1}{\sqrt[4]{6k}}\left[(1+r)^{1/4}+\frac{1}{2}(1+r)^{5/4} \right].
\label{eq:critkvk}
\end{equation}
$\epsilon_{c}$ is therefore the highest pumping value for which we may use the linearised fluctuation analysis~\cite{KVK3}.  In this expression, $r=\gamma_{b}/\gamma_{a}$ and $k = \kappa^{2}/9\gamma_{a}\gamma_{b}$.  For the parameters we use here, $\gamma_{a}=1$, $\gamma_{b}=2\gamma_{a}$, $\kappa=10^{-3}$, $\epsilon_{c}\approx 137$. Above this value, the system exhibits self-pulsing behaviour as in second harmonic generation~\cite{SHGpulse}. This behaviour is shown in Fig.~\ref{fig:THGpulse}, with results from both the classical equations and the TPPA equations. The solutions given by Geverkyan \etal are close to our solutions of the classical equations with a complex initial condition, while the TPPA equations show damped oscillations. The critical point is similar to SHG, in that we have two of the eigenvalues with zero real part and conjugate imaginary parts, but dissimilar in that the classical equations do not calculate mean fields correctly above this. Below $\epsilon_{c}$, numerical integration of the classical equations gives identical solutions to the TPPA equations, so that we have used standard numerical integration to calculate the spectral results presented below.

\begin{figure}[tbhp]
\includegraphics[width=0.75\columnwidth]{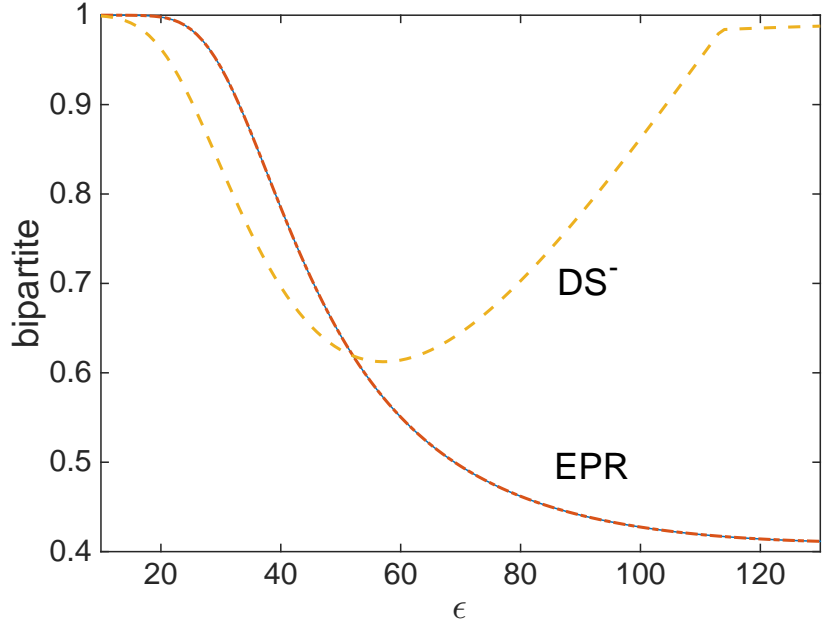}
\caption{(colour online) The minimum spectral values of the bipartite correlations at any frequency, as a function of cavity pumping, for the intracavity direct system. The parameters used are $\kappa=10^{-3}$,  $\gamma_{a}=1$ and $\gamma_{b}=2\gamma_{a}$.  Note that we have divided the Duan-Simon correlation by $4$ for simple comparison with the EPR values. For these parameters, $\epsilon_{c}\approx 137$.}
\label{fig:THGbicav}
\end{figure}

Fig.~\ref{fig:THGbicav} shows the minimum values of the EPR-steering and Duan-Simon correlations for the direct system, for pumping values below critical. Above the self-pulsing threshhold, there are no stationary solutions and the spectra cannot be obtained by the method used here. As the pumping amplitude is increased, the EPR value tends toward zero, while there is again less violation of the Duan-Simon inequality. Nevertheless, since steerable states are a subset of the entangled states violation of the Reid-EPR inequality shows that this system is a source of bright entangled modes with a large frequency difference. In the intracavity configuration, we find that the EPR-steering is totally symmetric.

\subsection{Intracavity cascaded conversion}
\label{subsec:cavcascade}

In order to calculate the spectra for this case, we proceed as in the linearised fluctuation analysis above. The equations for the mean fields are found by removing the noise terms from Eq.~\ref{eq:PPcascade}, leaving three coupled nonlinear differential equations. Analytic solutions are possible, but are not particularly illuminating~\cite{Yu3}. In what follows, we have calculated the steady-state solutions numerically. We find the drift matrix for the fluctuations as
\begin{equation}
A_{c} =
\begin{bmatrix}
\gamma_{0} & \kappa_{1}\alpha_{1} & \kappa_{1}\alpha_{0}^{\ast} & 0 & 0 & 0 \\
\kappa_{1}\alpha_{1}^{\ast} & \gamma_{0} & 0 & \kappa_{1}\alpha_{0} & 0 & 0 \\
-2\kappa_{1}\alpha_{0} & \kappa_{2}\alpha_{2} & \gamma_{1} & 0 & \kappa_{2}\alpha_{0}^{\ast} & 0 \\
 \kappa_{2}\alpha_{2}^{\ast} & -2\kappa_{1}\alpha_{0}^{\ast} & \gamma_{1} & 0 & \kappa_{2}\alpha_{0} \\
 -\kappa_{2}\alpha_{1} & 0 & -\kappa_{2}\alpha_{0} & 0 & \gamma_{2} & 0 \\
 0 & -\kappa_{2}\alpha_{1}^{\ast} & 0 &- \kappa_{2}\alpha_{0}^{\ast} & 0 & \gamma_{2}
\end{bmatrix},
\label{eq:Acmat}
\end{equation}
and the diffusion matrix
\begin{equation}
D_{c} = 
\begin{bmatrix}
-2\kappa_{1}\alpha_{1} & 0 & -\kappa_{2}\alpha_{2} & 0 & 0 & 0 \\
0 & -2\kappa_{1}\alpha_{1}^{\ast} & 0 & -\kappa_{2}\alpha_{2}^{\ast} & 0 & 0 \\
-\kappa_{2}\alpha_{2} & 0 & 0 & 0 & 0 & 0 \\
0 & -\kappa_{2}\alpha_{2}^{\ast} & 0 & 0 & 0 & 0 \\
0 & 0 & 0 & 0 & 0 & 0 \\
0 & 0 & 0 & 0 & 0 & 0
\end{bmatrix}.
\label{eq:Dcmat}
\end{equation}

\begin{figure}[tbhp]
\includegraphics[width=0.75\columnwidth]{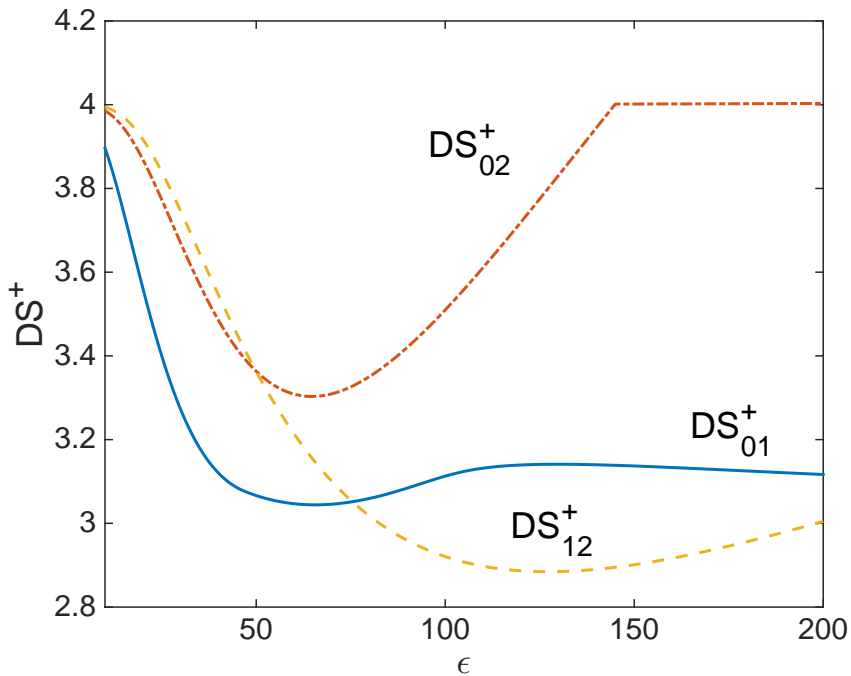}
\caption{(colour online) The minimum spectral values of the output Duan-Simon correlations at any frequency, as a function of cavity pumping, for the intracavity cascaded system. The nonlinearities are as $\kappa_{1}=10^{-2}$ and $\kappa_{2}=1.5\kappa_{1}$. The cavity loss rates are $\gamma_{1}=0.75\gamma_{0}$ and $\gamma_{2}=1.25\gamma_{0}$, with $\gamma_{0}=1$.}
\label{fig:DScavcascade}
\end{figure}

\begin{figure}[tbhp]
\includegraphics[width=0.75\columnwidth]{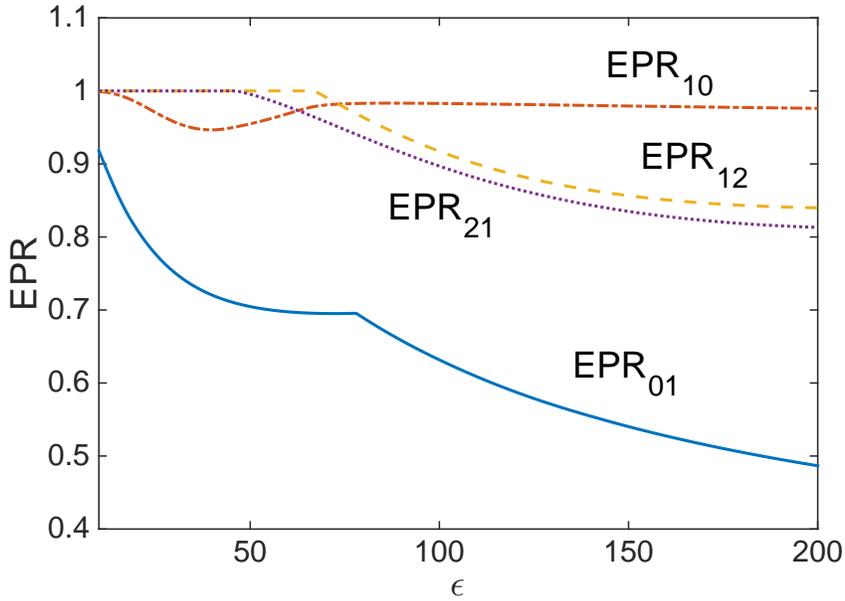}
\caption{(colour online) The minimum spectral values of the output Reid-EPR correlations at any frequency, as a function of cavity pumping, for the intracavity cascaded system. The nonlinearities are as $\kappa_{1}=10^{-2}$ and $\kappa_{2}=1.5\kappa_{1}$. The cavity loss rates are $\gamma_{1}=0.75\gamma_{0}$ and $\gamma_{2}=1.25\gamma_{0}$, with $\gamma_{0}=1$.}
\label{fig:EPRcavcascade}
\end{figure}

In Fig.~\ref{fig:DScavcascade} we show the minimum values of the Duan-Simon criteria at any frequency, as the pumping is increased from $10\gamma_{0}$ to $200\gamma_{0}$. The eigenvalues of the drift matrix $A_{c}$ had fully positive real parts over the parameter range shown, although they can develop negative real parts for other parameters~\cite{YuWang}, in particular when $\kappa_{2}\gg\kappa_{1}$. We have chosen our parameters to obtain comparable intensities in each field, which is not the case for that ratio of $\kappa_{2}/\kappa_{1}$. What is visible from the figure is that the Duan-Simon criteria find bipartite entanglement over the whole of this range of pumping for the bipartitions $01$ and $12$, but not for $02$. This is consistent with the results of Fig.~\ref{fig:EPRcavcascade} for the Reid-EPR correlation, where no violation of the inequality was found for the bipartition $02$. We also see that there is a small degree of asymmetric steering~\cite{Wisesteer} over some pumping range for the bipartition $12$, although this is possibly not experimentally significant once added sources of experimental noise have been taken into account. 

This system therefore allows for bipartite entanglement among three fields at different frequencies, although only entanglement and not EPR-steering were found between the fundamental and the third harmonic. We found no evidence of tripartite entanglement or inseparability as indicated by the van Loock-Furusawa inequalities~\cite{vLF}. The presence two different entangled bipartitions, with only one where EPR-steering is present can allow for some flexibility in quantum key distribution, as previously analysed for a system combining down-conversion with sum frequency generation~\cite{promiscuity}.

\section{Conclusion}

We have analysed two different quantum optical systems which can produce output light at three times the input pumping frequency, in terms of both their mean-field behaviour and entanglement properties. We have analysed both idealised travelling wave models and more realistic intracavity configurations. The direct system of third harmonic generation was analysed using a truncated positive-P approximation, whose accuracy has previously been shown in the travelling wave approximation. This was necessary because the full generalised equations present severe stability problems, and is expected to be more accurate for the intracavity configuration. The cascaded system, combining second harmonic generation with sum frequency generation, can be analysed exactly using the positive-P representation.

We have found that some previously presented results for the intensities of the fields in certain parameter regimes were not accurate, and explained this by comparison of semi-classical and quantum solutions. In particular, the self-pulsing previously predicted for direct third harmonic generation does not exhibit oscillations as large as previously calculated, and the steady-state of the intensities predicted for the cascaded process did not survive a quantum analysis. This is reminiscent of previous cases where classical and quantum mean-field predictions have been found to be completely different, for example with the revival of oscillations in the anharmonic oscillator and revivals of the fundamental in single-pass second harmonic generation. 

Both systems have been shown to produce output modes which exhibit bipartite entanglement and EPR-steering. Direct conversion shows a larger degree of violation of the EPR-steering inequalities while both configurations demonstrate asymmetric steering in some parameter regimes. The cascaded system provides three outputs at different frequencies, which could provide for flexibility in any practical applications.

\section*{Acknowledgments}

This research was supported by the Australian Research Council under the Future Fellowships Program (Grant ID: FT100100515).



\end{document}